\documentclass[10pt,journal,final,finalsubmission,twocolumn]{IEEEtran}

\usepackage{stmaryrd}
\usepackage{bm}
\usepackage{mathrsfs}
\usepackage{textcomp}
\usepackage{epsfig}
\usepackage{indentfirst}
\usepackage{amsmath}
\usepackage{amssymb}
\usepackage{subfigure}
\usepackage{CJK}
\usepackage{amsthm}
\usepackage{multicol}
\usepackage{lipsum}
\usepackage{arydshln}
\usepackage{color}

\usepackage{cases}

\usepackage{algorithm}
\usepackage{algorithmic}

\usepackage{epstopdf}

\theoremstyle{plain}

\theoremstyle{remark}

\usepackage{cite}

\begin{document}
\title{{LDPC-Based Code Hopping for Gaussian Wiretap Channel With Limited Feedback}}
\author{
{Zhao Chen$^{1,3}$, Liuguo Yin$^{2,3}$, and Jianhua Lu$^1$}\\
State Key Laboratory on Microwave and Digital Communications \\
Tsinghua National Laboratory for Information Science and Technology \\
$^1$Department of Electronic Engineering, Tsinghua University\\ $^2$School of Aerospace, Tsinghua University\\
$^3$EDA Lab, Research Institute of Tsinghua University in Shenzhen\\
Email: zhao-chen10@mails.tsinghua.edu.cn; \{yinlg, lhh-dee\}@tsinghua.edu.cn
}
\maketitle

\begin{abstract}
This paper presents a scheme named code hopping (CodeHop) for gaussian wiretap channels based on nonsystematic low-density parity-check (LDPC) codes.
Different from traditional communications, in the CodeHop scheme, the legitimate receiver (Bob) rapidly switches the parity-check matrix  upon each correctly received source message block.
Given an authenticated public feedback channel, the transmitter's (Alice) parity-check matrix can also be synchronized with the receiver's.
As a result, once an eavesdropper (Eve) erroneously decodes a message block, she may not be able to follow the update of subsequent parity-check matrices.
Thus, the average BER of Eve will be very close to $0.5$ if the transmitted number of message blocks is large enough.
Focused on the measure of security gap defined as the difference of channel quality between Bob and Eve, numerical results show that the CodeHop scheme outperforms other solutions by sufficiently reducing the security gap without sacrificing the error-correcting performance of Bob.


\end{abstract}


\section{Introduction}\label{sec.Intro}
Traditionally, security of data transmissions has been handled at upper protocol layers, which presents wiretappers (Eve) by cryptographic algorithms with a secret key privately shared between a transmitter (Alice) and legitimate receivers (Bob).
This notion of security relies on the assumption that Eve has limited resources.
Thus, \emph{perfect secrecy} was introduced by Shannon~\cite{shannon1949communication}, which proved that if the secret key rate is lager or equal to the transmission rate, information-theoretical security can be achieved such that the encrypted message reveals no information of the source message.

Owing to the stochastic nature of communication channels, the work of Shannon was extended to the model of \emph{wiretap channel} by Wyner~\cite{wyner1975wire}, where information-theoretically secure communication is possible when Eve's channel condition is worse than Bob's, and \emph{secrecy capacity} is measured as the highest transmission rate that can be achieved by Bob without any information leaked to Eve.
Hereafter, Wyner's work was refined by Csiszar and Korner in \cite{csiszar1978broadcast}, and then generalized to gaussian wiretap channel in \cite{leung1978gaussian}.
Note that the equivocation rate at Eve can also be a metric for security, which is defined as the conditional entropy of the source message with given Eve's noisy observation.
More recently, scenarios including fading channels~\cite{gopala2008secrecy}, MIMO channels~\cite{oggier2011secrecy} and multi-user channels~\cite{ekrem2011secrecy} have also been investigated in the literature.

Recently, coding techniques attract much research interests to achieve better security performance. 
In~\cite{thangaraj2007applications}, for binary erasure channel (BEC) and  binary symmetric channel (BSC), low-density parity-check (LDPC)-based wiretap codes are constructed.
For gaussian wiretap channels, lattice codes are also developed in~\cite{oggier2011lattice}.
Moreover, with the emerging polar codes techniques, a nested code structure is established for binary-input symmetric-output (BISO) channels in \cite{koyluoglu2012polar}.
All the designs above are for asymptotically large block lengths, which may be difficult to implement in practical systems.
To this end, an LDPC code design for BPSK-constrained gaussian wiretap channel was shown in~\cite{wong2011ldpc}. A joint error correction and encryption using LDPC codes is also studied in~\cite{chen2016codehop}.


Recently, in order to construct practical codes, the bit error rate (BER) of message bits is measured for physical layer security~\cite{chen2015hamming}.
Specifically, if Eve's BER is ensured to be close to $0.5$ and the error bits are i.i.d. distributed, she would be unable to recover any information from the decoded messages.
Then, \emph{security gap} was introduced by Klinc~\cite{klinc2009ldpc}, which is defined as the difference of the channel quality between Bob and Eve, requiring Eve's BER to be sufficiently secure, while keeping Bob's BER low enough to reliably receive messages.
In \cite{baldi2012coding}, Baldi reduced the security gap by implementing nonsystematic LDPC codes using scrambling matrices, where security is enhanced by considering automatic repeat-request (ARQ).
Also, a feedback-aided secure transmission scheme using stopping sets of LDPC codes can be found in \cite{harrison2011coding}.

In this paper, the gaussian wiretap channel enhanced with an authenticated feedback channel is considered, where an LDPC-based code hopping (CodeHop) scheme is proposed to improve physical layer security by reducing the security gap.
Using a structured-random LDPC code generator, the parity-check matrix of CodeHop will be rapidly updated upon a correctly received message block by Bob, which is then notified to Alice with one bit feedback.
Once Eve encounters a decoding error, she may not be able to decode any more message block without knowing the parity-check matrix in the following transmission and then her average BER will be close to $0.5$.
Thus, the security gap can be significantly reduced.

The rest of the paper is organized as follows. In Section II, system model of the CodeHop scheme is introduced. Then, the construction of the structured-random LDPC codes is shown in Section III. In Section IV, the security gap performance is analyzed. Finally, we conclude this paper in Section V.

\section{The LDPC-Based Code Hopping Scheme}\label{sec.CodeHop}
In this section, system model of the gaussian wiretap channel with limited feedback will be introduced, along with the design of the encoder and the decoder.

\subsection{System Model}\label{sec.Model}
\begin{figure}
\centering
\includegraphics[width=9cm]{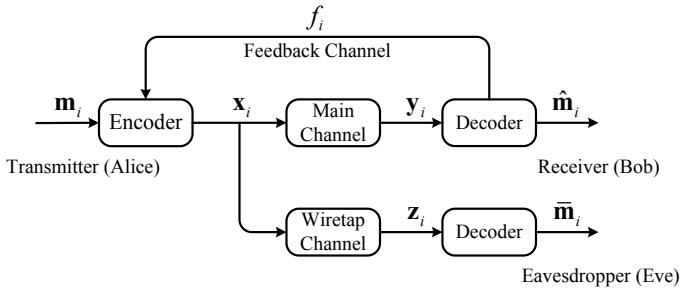}
\caption{System model of the gaussian wiretap channel with limited feedback.}
\label{fig.wiretap}
\end{figure}

As depicted in Fig. \ref{fig.wiretap}, for $i = 1,2,\ldots$, the secret source to be transmitted is divided into $k$-bit message blocks $\mathbf{m}_i$, which are encoded into $n$-bit codewords $\mathbf{x}_i$ by the encoder at Alice and then transmitted over an AWGN channel to Bob.
In the meantime, codewords are also perceived by Eve over a noisier and independent AWGN channel.
For Bob and Eve, the received codewords are denoted by $\mathbf{y}_i$ and $\mathbf{z}_i$, which are recovered by the decoders as $\hat{\mathbf{m}}_i$ and $\bar{\mathbf{m}}_i$, respectively.
Additionally, in our model, legitimate receivers are always given a public feedback channel, which can be used to inform Alice whether the current codeword is correctly decoded with just one bit feedback $f_i$.

Let $P^{B}_e$ and $P^{E}_e$ define the receiving BERs of Bob and Eve, respectively.
To guarantee both information reliability and security, it is desired to constrain $P^{B}_e$ to be lower than a fixed threshold $P_{e,\max}^B(\thickapprox0)$ and $P^{E}_e$ to be greater than a threshold $P_{e,\min}^E(\thickapprox0.5)$.
If the receiving SNRs at Bob and Eve are denoted by $\mathrm{SNR}_{B}$ and $\mathrm{SNR}_{E}$, it must hold that
\begin{numcases}{}
        \mathrm{SNR}_{B} \geq \mathrm{SNR}_{B,\min} = p^{-1}(P_{e,\max}^B), \label{eq.bob_e}\\
        \mathrm{SNR}_{E} \leq \mathrm{SNR}_{E,\max} = p^{-1}(P_{e,\min}^E), \label{eq.eve_e}
\end{numcases}
where $p(\cdot)$ denotes the BER as a function of the SNR, and the inverse $p^{-1}(\cdot)$ gives the target SNR for a given BER threshold.
Then, the security gap is defined as
\begin{equation}\label{eq.gap}
  S_g = \mathrm{SNR}_{B,\min} - \mathrm{SNR}_{E,\max},
\end{equation}
where the SNRs are expressed in decibels (dB).

If both \eqref{eq.bob_e} and \eqref{eq.eve_e} are satisfied, the secrecy rate $R_s$ can be defined as the efficient transmission rate
\begin{equation}\label{eq.secrecy_rate}
  R_s \thickapprox R = \frac{k}{n},
\end{equation}
where in fact $R_s$ is slightly less than $R$ due to some overhead bits in $\mathbf{m}_i$ for integrity check.
Note that the security gap $S_g$ measures the minimum required difference between Bob and Eve's SNR in dB for secure communication, and our target is to reduce the security gap as small as possible.

\subsection{Encoder and Decoder Design}

\begin{figure}
\centering
\subfigure[Encoder.]
{
    \label{fig.encoder}
    \includegraphics[width=5.5cm]{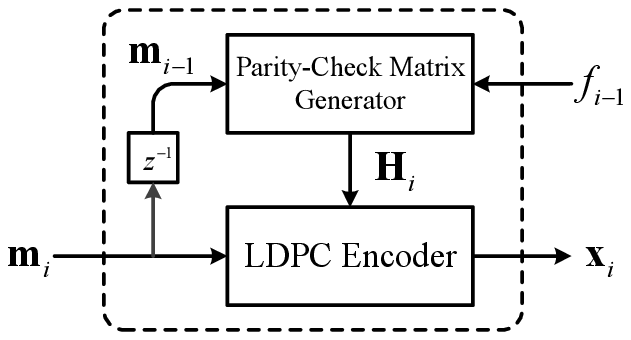}
}

\subfigure[Decoder.]
{
    \label{fig.decoder}
    \includegraphics[width=6.5cm]{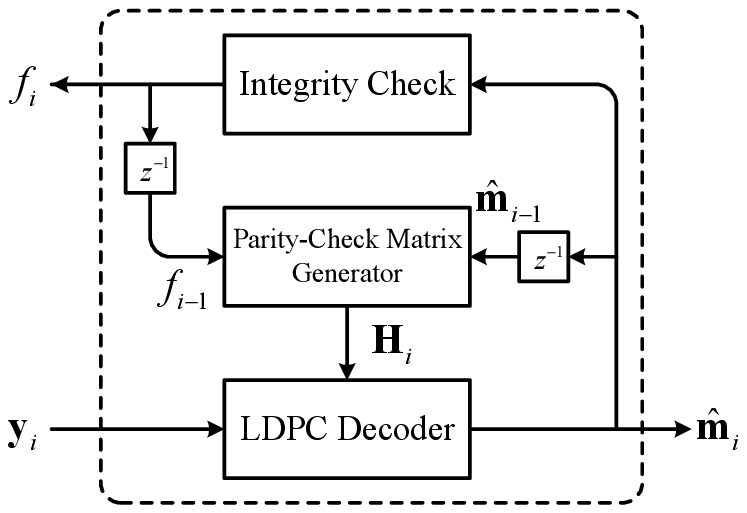}
}
\caption{Block diagrams of the encoder and the decoder. Note that the $z^{-1}$ block denotes a delay unit.}
\label{Fig.design} 
\end{figure}

In order to minimize the security gap of gaussian wiretap channel, the CodeHop scheme is implemented such that each source message block $\mathbf{m}_i$ will be encoded by a rapidly hopping parity-check matrix of LDPC codes $\mathbf{H}_i$.
That is, $\mathbf{H}_i$ is not fixed for all codewords. On the contrary, $\mathbf{H}_i$ will be generated in real time according to the past correctly received source message block $\mathbf{m}_{i-1}$ by a parity-check matrix generator.
Block diagrams of the encoder and the decoder are both illustrated in Fig. \ref{Fig.design}.

In the encoder of Alice, starting from a public agreed parity-check matrix $\mathbf{H}_0$ and feedback $f_0 = 0$, the LDPC parity-check matrix $\mathbf{H}_i$ is generated by
\begin{equation}\label{eq.hash}
    \mathbf{H}_i =
\begin{cases}
        \mathbf{H}(hash(\mathbf{m}_{i-1})), & \hbox{if $f_{i-1} = 1$;} \\
        \mathbf{H}_{i-1}, & \hbox{if $f_{i-1} = 0$;}
    \end{cases}
\end{equation}
where $hash(\cdot)$ is a hash function, and $\mathbf{H}(\cdot)$ is the mapping of structured-random LDPC codes described in Section \ref{sec.randomLDPC}.
Thus, the parity-check matrix $\mathbf{H}_i$ will be updated when the past source message $\mathbf{m}_{i-1}$ is successfully recovered by Bob, i.e. $f_{i-1} = 1$.
Otherwise, $\mathbf{H}_i$ will be kept the same as $\mathbf{H}_{i-1}$.

In the decoder of Bob, the public feedback $f_i$ is the result of integrity check for $\mathbf{\hat{m}}_i$.
Although the correctness of $\mathbf{m}_i$ can be determined by the syndrome of the decoded codeword in most cases, it is still possible that there exist some undetected errors in $\mathbf{\hat{m}}_i$ when the decoded codeword converges to another valid codeword of $\mathbf{H}_i$.
Thus, integrity check, e.g. cyclic redundancy check (CRC), is still needed to avoid such undetected errors, and some overhead bits are required in $\mathbf{m}_i$ to store the integrity check value.
As for the parity-check matrix $\mathbf{H}_i$, it will be generated as in the encoder of Alice.


\section{Structured-Random LDPC Codes}\label{sec.randomLDPC}
In this section, we will show how to generate nonsystematic LDPC codes in real time via structured-random protograph expanding.
As presented by Thorpe in~\cite{thorpe2003low}, protograph can be regarded as the minimal base Tanner graph to describe an LDPC code.
Using protograph, our target is to puncture all of the $k$ information bits and transmit only the $n$ parity bits in the original $(n+k)$-bit codeword.
In order to avoid stopping sets which make the iterative decoding of nonsystematic codes difficult to converge, the code doping method~\cite{ten2003design} is adopted.


%

\begin{figure}
\centering
\includegraphics[width=6cm]{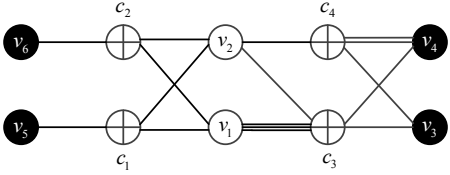}
\caption{Protograph $P$ optimized for the rate-$1/2$ nonsystematic LDPC code.}
\label{fig:protograph}
\end{figure}

As for a rate-$1/2$ nonsystematic LDPC code, the result of an optimized protograph $P=(V,C,E)$ is given in Fig.~\ref{fig:protograph},
which is equivalent to a $4\times6$ base parity-check matrix $\mathbf{H}_{\mathrm{B},0}$ as shown below,
\begin{align}
    {\mathbf{H}_{\mathrm{B},0}} & = \left[
 \begin{array}{cc;{2pt/2pt}cccc}
 1 & 1 & 0 & 0 & 1 & 0 \\
 1 & 1 & 0 & 0 & 0 & 1 \\
 3 & 1 & 1 & 1 & 0 & 0 \\
 0 & 1 & 1 & 2 & 0 & 0
\end{array}\right].\label{eq:base}
\end{align}
In the figure, there are a variable node set $V=\{v_1, v_2, ..., v_6\}$, a check node set $C=\{c_1, c_2, ..., c_4\}$ and an edge set $E=\{e_1, e_2, ..., e_{16}\}$ in the protograph $P$.
Among all variable nodes, the information nodes denoted by $v_1$ and $v_2$ will be punctured.
Thus, according to the requirement of code doping, the check node $c_4$ is designated to be connected to only one punctured variable node $v_2$ to trigger the convergence of the iterative decoding.

Starting with the protograph $P$, a larger tanner graph can be obtained by copy-and-permute operation.
Let each edge $e \in E$ stand for a unique edge type.
If expanded with a factor of $T$, the copy-and-permute operation firstly replicates the protograph $P$ for $T$ times and forms a set of $T$ edge copies for each edge type, whose endpoints are then permuted among the variable and check nodes in the set. Thus the $T$ copies of the protograph are all interconnected in the derived graph, which defines the Tanner graph for the derived code.
Note that the operation of edge expanding is equivalent to matrix expanding for each element in the base matrix $\mathbf{H}_{\mathrm{B},0}$.
Each element of value $w$ will be expanded to a $T\times T$ matrix with $w$ ones in each row or column, and thus the equivalent parity-check matrix can be of size $4T\times6T$.

Instead of using random permutations on the set of each edge type, permutations defined by algebraic structures such as cyclic shift are preferred for the ease of description and efficient implementation.
That is to say, the protograph can be expanded by choosing appropriate $T \times T$ circulant permutation matrices $\mathbf{I}_{T}(t)$ for each edge type, and the derived parity-check matrix will become a $T$-circulant matrix.
Note that each row in $\mathbf{I}_{T}(t)$ can be obtained by one bit cyclic right shifting its previous row, and thus $\mathbf{I}_{T}(t)$ is defined by its first row $\mathbf{u}_{T}(t)$, where $\mathbf{u}_{T}(t) = (0,\ldots,0,1,0,\ldots,0)$ and the shift value $t \in \llbracket 0,T-1 \rrbracket$~\footnote{Note that $\llbracket a,b \rrbracket :=\{a,a+1,\ldots,b\} $, which denotes the set of integers between $a$ and $b$.} denotes the position of the unique one in the vector.
Therefore, given the protograph $P$, the derived LDPC code can be described by shift values of all the circulant permutations.

Usually searching for only one good code is satisfactory for channel coding. However, now it is far more challenging to construct a large amount of highly efficient LDPC codes for the CodeHop scheme.
Thus, expanding with just one single stage is not enough.
A multi-stage expanding scheme is developed, which we referred to as a structured-random protograph expanding technique.
Specifically, the protograph $P$ will be expanded via $L > 1$ stages such that the total expanding factor $T = T_1 T_2 \cdots T_L$, where $T_l$ is the expanding factor of the $l$th stage for $l \in \llbracket 1,L \rrbracket$ and the equivalent expanded parity-check matrix is denoted by $\mathbf{H}_{\mathrm{B},l}$.
\begin{itemize}
  \item \emph{Structured expanding}: In the first $L-1$ stages, the protograph is carefully expanded to avoid short loops, low-weight codewords or parallel edges, which can restrict the general structure of the code. After $L-1$ stages, all the non-zero elements in $\mathbf{H}_{\mathrm{B},{L-1}}$ will equal $1$.
  \item \emph{Random expanding}: In the $L$th stage, all of the edges in the tanner graph will be randomly expanded, or equivalently, all the non-zero elements in $\mathbf{H}_{\mathrm{B},{L-1}}$ will be expanded to circulant permutations according to the hash value of the past correctly decoded message $\mathbf{m}_{i-1}$.
\end{itemize}


After expanding the protograph $P$ in Fig.~\ref{fig:protograph}, each parity-check matrix $\mathbf{H}_i$ is a $T_L$-circulant matrix with a size of ${n \times (n+k)}$, which can be written as $\mathbf{H}_i = \left[ {{\mathbf{A}},{\mathbf{B}}} \right]$ such that
\begin{align}
    {\mathbf{A}} & = \begin{bmatrix}\mathbf{A}_{\alpha\beta}^{w} \end{bmatrix}_{2 \times 4} =
 \begin{bmatrix}
{{\mathbf{A}}_{11}^1}&{{\mathbf{A}}_{12}^1}\\
{{\mathbf{A}}_{21}^1}&{{\mathbf{A}}_{22}^1}\\
{{\mathbf{A}}_{31}^3}&{{\mathbf{A}}_{32}^1}\\
{\mathbf{0}}&{{\mathbf{A}}_{42}^1}
\end{bmatrix},\label{eq:a} \\
{\mathbf{B}} & = \begin{bmatrix}\mathbf{B}_{\alpha\beta}^{w} \end{bmatrix}_{4 \times 4} =
 \begin{bmatrix}
{\mathbf{0}}&{\mathbf{0}}&{{\mathbf{B}}_{13}^1}&{\mathbf{0}}\\
{\mathbf{0}}&{\mathbf{0}}&{\mathbf{0}}&{{\mathbf{B}}_{24}^1}\\
{{\mathbf{B}}_{31}^1}&{{\mathbf{B}}_{32}^1}&{\mathbf{0}}&{\mathbf{0}}\\
{{\mathbf{B}}_{41}^1}&{{\mathbf{B}}_{42}^2}&{\mathbf{0}}&{\mathbf{0}}
\end{bmatrix}. \label{eq:b}
\end{align}
Note that $n = 4T$, $k = 2T$, and the first $k=2T$ nodes are punctured as information nodes among all the  $n+k=6T$ variable nodes.
There are totally $J= |E| T/T_L$ non-zero elements in $\mathbf{H}_{\mathrm{B},L-1}$, each of which will be then replaced by a $T_L\times T_L$ circulant permutation matrix $\mathbf{I}_{T_L}(t)$.
Meanwhile, all the zero elements in $\mathbf{H}_{\mathrm{B},L-1}$ will be replaced by a $T_L\times T_L$ zero matrix $\mathbf{0}_{T_L\times T_L}$.
Now, we rewrite the hash value $hash(\mathbf{m}_{i+1})$ as a binary vector
\begin{equation}\label{eq:ri}
    \mathbf{r}_i = hash(\mathbf{m}_{i-1}) = (r_{i,0},r_{i,1},\ldots,r_{i,j},\ldots,r_{i,J-1}),
\end{equation}
where each number $r_{i,j} \in \llbracket 0,T_L-1 \rrbracket$ is represented by $\log_{2}{T_L}$ bits in the vector $\mathbf{r}_i$.
Therefore, all the random shift values in the $L$th stage will be controlled by the vector $\mathbf{r}_i$, whose length is required to be $h = J \log_{2}{T_L}$ bits.
If the bit length of $hash(\mathbf{m}_{i-1})$ is smaller than $h$, $\mathbf{r}_i$ can be filled up by using $hash(\mathbf{m}_{i-1})$ repeatedly.
As for the parameters selected in the first $L-1$ stages, i.e. all the expanding factors and shift values, they are constant and will be publicly sent to Bob in advance.

\begin{figure}
\centering
\includegraphics[height=6cm]{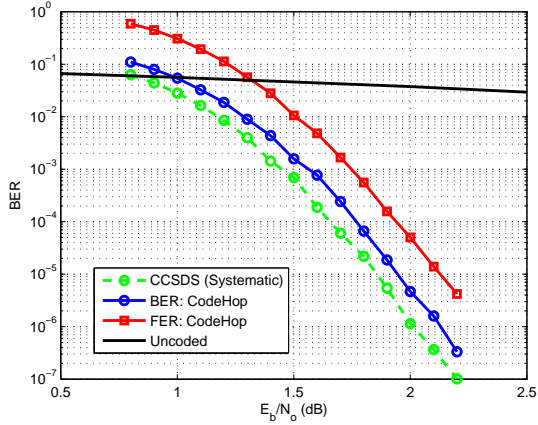}
\caption{For $(2048,1024)$ LDPC codes, the average BER and FER of the nonsystematic LDPC codes in CodeHop is compared with the BER of the systematic code specified in CCSDS.}
\label{fig:ber}
\vspace{-1em}
\end{figure}

\subsection{An example}\label{sec.ex}
With the structured-random expanding technique, a hopping sequence of parity-check matrices for $(2048,1024)$ nonsystematic LDPC codes is generated.
From the protograph $P$ in Fig.~\ref{fig:protograph}, each parity-check matrix $\mathbf{H}_i$ will be constructed by a total expanding factor $T = T_1 T_2 T_3 = 4 \times 4 \times 32 = 512$ via $L=3$ stages.
In the first two stages, the protograph $P$ is expanded twice by factors of $T_1=T_2=4$, which is intended to separate all the parallel edges and guarantees the girth is larger than $4$ in the derived Tanner graph.
In the third stage, totally there are $|E| T/T_3 = 256$ edges to be randomly expanded by a factor of $T_3=32$, which is described by a vector $\mathbf{r}_i$ with length {of $h =J \log_{2}{T_L}= 1280$ bits}.
If SHA-256 is adopted as the $hash(\cdot)$, the $256$-bit hash value $hash(\mathbf{m}_{i-1})$ will be repeated by four times to represent $\mathbf{r}_i$.
Thus, a large set of parity-check matrices $\mathcal{H} = \{\mathbf{H}(\mathbf{r}):\mathbf{r} \in \llbracket 0,2^{256}-1 \rrbracket\}$ is available to be randomly generated in the hopping sequence of CodeHop.

To evaluate the error correcting performance of CodeHop, it is different from the evaluation of a fixed LDPC parity-check matrix.
The BER of one hopping sequence will be evaluated as how the CodeHop scheme works, that is, by performing numerical simulations with rapidly switching parity-check matrices.
For the average performance of CodeHop, random initial $\mathbf{H}_0$ will be tested to generate different hopping sequences, whose BERs and FERs are averaged as the final result.
As shown in Fig.~\ref{fig:ber}, the average BER and FER of the structured-random nonsystematic $(2048,1024)$ LDPC codes are plotted.
The number of iterations is restricted by $63$.
It can be seen that the average BER of nonsystematic LDPC codes is slightly degraded by no more than $0.2$ dB in $E_b / N_0$, compared with the systematic codes specified in CCSDS.

\section{Security Gap Analysis}
In this section, the analysis of security gap using CodeHop for gaussian wiretap channel with limited feedback will be presented.
Owing to the public feedback $f_i$, the parity-check matrix $\mathbf{H}_i$ will be always perfectly synchronized between Alice and Bob.
However, Eve is restricted to passively receive the feedback $f_i$.
As a result, an event which we referred to as \emph{synchronization error} may happen.
That is, there exists an index $i_{\mathrm{TH}} \in \mathbb{N}$, such that $\mathbf{z}_{i_{\mathrm{TH}}}$ is not correctly decoded by Eve, but $\mathbf{y}_{i_{\mathrm{TH}}}$ is successfully recovered by Bob.
Thus, Eve will be not able to follow the update of the parity-check matrix $\mathbf{H}_i$ for $i \geq i_{\mathrm{TH}}$, and the subsequent source message blocks $\mathbf{m}_i$ cannot be decoded by Eve without any knowledge of $\mathbf{H}_i$.

Let $\mathrm{SNR}_B$ and $\mathrm{SNR}_E$ denote the receiving SNRs of Bob and Eve, respectively. The BER of Bob is given by
\begin{equation}\label{eq.bob}
  P_e^{B} = P_e(\mathrm{SNR}_B),
\end{equation}
where $P_e(\mathrm{SNR})$ is the BER of the CodeHop scheme with perfect synchronization that can be obtained from Fig. \ref{fig:ber}.

As for Eve's BER, firstly the event of synchronization error needs to be investigated.
Assume that the total number of source message blocks is $N$.
Given $P_f(\mathrm{SNR})$ as the frame error probability (FER) of the CodeHop scheme with perfect synchronization, the index $i_{\mathrm{TH}}$ of the first erroneously synchronized message block follows a geometric distribution.
Thus, for $1\leq i_{0} \leq N$, the probability of the event $i_{\mathrm{TH}} = i_{0}$ can be derived by
\begin{equation}\label{eq.th}
  \mathrm{Pr}(i_{\mathrm{TH}} = i_0) = \left[1-(1-P_f^B)P_f^E\right]^{i_0-1}(1-P_f^B)P_f^E,
\end{equation}
 where we define $P_f^B = P_f(\mathrm{SNR}_B)$ and $P_f^E = P_f(\mathrm{SNR}_E)$.
If synchronization error never happens, the probability is denoted by
\begin{equation}\label{eq.th_n_plus_1}
  \mathrm{Pr}(i_{\mathrm{TH}} \geq N+1) = [1-(1-P_f^B)P_f^E]^N.
\end{equation}
Then, for a fixed index $i_{\mathrm{TH}}$, Eve's erroneously received source message blocks can be divided into two categories.
The first category involves the blocks $\mathbf{m}_i$ for $1 \leq i \leq i_{\mathrm{TH}}-1$, which are erroneously decoded with knowing the parity-check matrix $\mathbf{H}_i$.
Actually, the number of block errors in the first category follows a binomial distribution and can be estimated by the mean value
\begin{equation}\label{eq.errNum_mean}
  N_{\mathrm{ER}}(i_{\mathrm{TH}}) = \frac{(i_{\mathrm{TH}}-1)P_f^B P_f^E}{1-(1-P_f^B)P_f^E},
\end{equation}
where the averaged number of bit errors in such an erroneous message block is given by
\begin{equation}\label{eq.average}
  k_{\mathrm{ER}} = \frac{P_e(\mathrm{SNR}_E)\cdot k}{P_f(\mathrm{SNR}_E)}.
\end{equation}
The second category includes all the unknown message blocks for $i_{\mathrm{TH}} \leq i \leq N$, which is caused by synchronization error and cannot be decoded without knowing $\mathbf{H}_i$.
That is, all of the message bits in $\mathbf{m}_i$ are randomly decoded and half of them will be error bits.
Finally, the BER of Eve can be obtained by
\begin{align}\label{eq.eve}
  {P}_e^{E} = \sum_{i=1}^{N} &\frac{k_{\mathrm{ER}} \cdot N_{\mathrm{ER}}(i) + 0.5k \cdot(N - i + 1)}{k\cdot N}\cdot \mathrm{Pr}(i_{\mathrm{TH}} = i) \nonumber \\ & + \frac{k_{\mathrm{ER}} \cdot N_{\mathrm{ER}}(N+1)}{k\cdot N}\cdot \mathrm{Pr}(i_{\mathrm{TH}} \geq N+1).
\end{align}

\begin{figure}
\centering
\includegraphics[height=6cm]{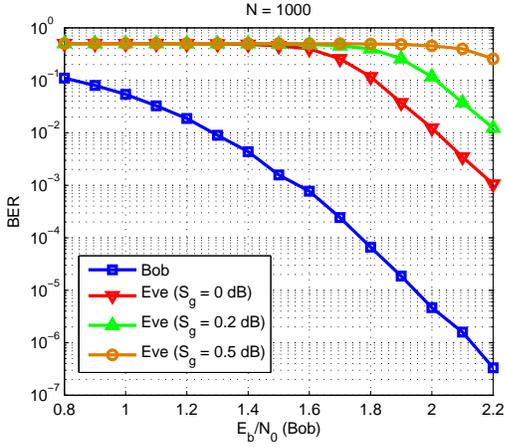}
\vspace{-0.5em}
\caption{BER of the CodeHop scheme for gaussian wiretap channel using $(2048,1024)$ nonsystematic LDPC codes for different security gaps when $N = 1000$ message blocks are transmitted.}
\label{fig:n_1000}
\end{figure}

\begin{figure}
\centering
\includegraphics[height=6cm]{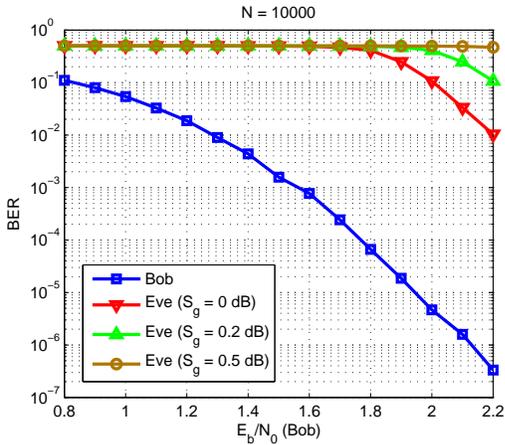}
\vspace{-0.5em}
\caption{BER of the CodeHop scheme for gaussian wiretap channel using $(2048,1024)$ nonsystematic LDPC codes for different security gaps when $N = 10000$ message blocks are transmitted.}
\label{fig:n_10000}
\vspace{-1em}
\end{figure}

From \eqref{eq.bob} and \eqref{eq.eve}, we can characterize the security gap of the CodeHop scheme for gaussian wiretap channel.
As shown in Fig. \ref{fig:n_1000} and Fig. \ref{fig:n_10000}, the BER curves for different security gaps $S_g$ and different number of transmitted message block $N$ using $(2048,1024)$ nonsystematic LDPC codes are illustrated.
If the BER threshold of Eve is set to be $P_{e,\min}^E = 0.49$, for $N = 1000$, we can see in Fig. \ref{fig:n_1000} that $S_g = 0.5 \mathrm{dB}$ is achievable for the BER threshold of Bob $P_{e,\max}^B < 10^{-5}$.
If $P_{e,\max}^B$ is relaxed to be less than $10^{-3}$, $S_g = 0 \mathrm{dB}$ can be also achieved.
It can be inferred that with the help of public feedback $f_i$ and rapidly hopping of the parity-check matrix $\mathbf{H}_i$, the security gap can be significantly reduced.
Moreover, if $N = 10000$, i.e. more source message blocks are transmitted, the security gap performance can be further improved.
E.g., for $P_{e,\max}^B < 10^{-5}$, the security gap $S_g$ can be reduced to $0.2 \mathrm{dB}$.
This is owed to the fact that all message blocks $\mathbf{m}_i$ after $i_{{\mathrm{TH}}}$ cannot be decoded by Eve.
Thus, more source message blocks to be transmitted makes Eve suffer from more random decoded blocks.
It is worth noting that to make the error distributed uniformly in the decoded message bits of Eve, interleaving is needed for all the $N$ transmitted source message blocks.

\vspace{-2pt}
\section{Conclusions}\label{sec.Conclusion}
Design of practical coding techniques for gaussian wiretap channel is investigated in this paper.
Based on nonsystematic LDPC codes, we propose the CodeHop scheme in which the LDPC parity-check matrix will be rapidly switched depending on the most recent successfully recovered message block at Bob.
Enhanced by an authenticated public feedback channel,
Alice can be notified of the decoding state of Bob and synchronize to the updated parity-check matrix with Bob.
However, Eve is restricted to passively follow the update of the parity-check matrix.
Once the event of synchronization error occurs, she would be unable to decode the rest of the message blocks.
In the CodeHop scheme, security gap is considered as a metric for physical layer security.
It is verified with numerical simulations that the security gap can be highly reduced with finite length LDPC codes.
In the future, more scenarios such as fading channels will also be studied by applying the CodeHop scheme.

\section*{Acknowledgment}
{This work was supported in part by the Natural Science Foundation of China (61101072, 61132002) and the new strategic industries development projects of Shenzhen city (ZDSY20120616141333842).}

\bibliographystyle{IEEEtran}
\bibliography{IEEEfull,CZBib}

\begin{thebibliography}{10}
\providecommand{\url}[1]{#1}
\csname url@rmstyle\endcsname
\providecommand{\newblock}{\relax}
\providecommand{\bibinfo}[2]{#2}
\providecommand\BIBentrySTDinterwordspacing{\spaceskip=0pt\relax}
\providecommand\BIBentryALTinterwordstretchfactor{4}
\providecommand\BIBentryALTinterwordspacing{\spaceskip=\fontdimen2\font plus
\BIBentryALTinterwordstretchfactor\fontdimen3\font minus
  \fontdimen4\font\relax}
\providecommand\BIBforeignlanguage[2]{{%
\expandafter\ifx\csname l@#1\endcsname\relax
\typeout{** WARNING: IEEEtran.bst: No hyphenation pattern has been}%
\typeout{** loaded for the language `#1'. Using the pattern for}%
\typeout{** the default language instead.}%
\else
\language=\csname l@#1\endcsname
\fi
#2}}

\bibitem{shannon1949communication}
C.~E. Shannon, ``Communication theory of secrecy systems,'' \emph{Bell system
  technical journal}, vol.~28, no.~4, pp. 656--715, 1949.

\bibitem{wyner1975wire}
A.~D. Wyner, ``The wire-tap channel,'' \emph{Bell system technical journal},
  vol.~54, pp. 1355--1387, 1975.

\bibitem{csiszar1978broadcast}
I.~Csisz{\'a}r and J.~Korner, ``Broadcast channels with confidential
  messages,'' \emph{IEEE Trans. Inf. Theory}, vol.~24, no.~3, pp. 339--348,
  1978.

\bibitem{leung1978gaussian}
S.~Leung-Yan-Cheong and M.~Hellman, ``The gaussian wire-tap channel,''
  \emph{IEEE Trans. Inf. Theory}, vol.~24, no.~4, pp. 451--456, 1978.

\bibitem{gopala2008secrecy}
P.~K. Gopala, L.~Lai, and H.~El~Gamal, ``On the secrecy capacity of fading
  channels,'' \emph{IEEE Trans. Inf. Theory}, vol.~54, no.~10, pp. 4687--4698,
  2008.

\bibitem{oggier2011secrecy}
F.~Oggier and B.~Hassibi, ``The secrecy capacity of the mimo wiretap channel,''
  \emph{IEEE Trans. Inf. Theory}, vol.~57, no.~8, pp. 4961--4972, 2011.

\bibitem{ekrem2011secrecy}
E.~Ekrem and S.~Ulukus, ``The secrecy capacity region of the gaussian mimo
  multi-receiver wiretap channel,'' \emph{IEEE Trans. Inf. Theory}, vol.~57,
  no.~4, pp. 2083--2114, 2011.

\bibitem{thangaraj2007applications}
A.~Thangaraj, S.~Dihidar, A.~R. Calderbank, S.~W. McLaughlin, and J.-M.
  Merolla, ``Applications of ldpc codes to the wiretap channel,'' \emph{IEEE
  Trans. Inf. Theory}, vol.~53, no.~8, pp. 2933--2945, 2007.

\bibitem{oggier2011lattice}
F.~Oggier, P.~Sol{\'e}, and J.-C. Belfiore, ``Lattice codes for the wiretap
  gaussian channel: Construction and analysis,'' \emph{arXiv preprint
  arXiv:1103.4086}, 2011.

\bibitem{koyluoglu2012polar}
O.~O. Koyluoglu and H.~El~Gamal, ``Polar coding for secure transmission and key
  agreement,'' \emph{IEEE Trans. Inf. Forensics Security}, vol.~7, no.~5, pp.
  1472--1483, 2012.

\bibitem{wong2011ldpc}
C.~W. Wong, T.~F. Wong, and J.~M. Shea, ``{LDPC} code design for the
  bpsk-constrained gaussian wiretap channel,'' in \emph{IEEE GLOBECOM
  Workshops}, 2011, pp. 898--902.

\bibitem{chen2016codehop}
Z.~Chen, L.~Yin, Y.~Pei, and J.~Lu, ``Codehop: physical layer error correction
  and encryption with ldpc-based code hopping,'' \emph{Science China
  Information Sciences}, vol.~59, no.~10, p. 102309, 2016.

\bibitem{chen2015hamming}
Z.~Chen, L.~Yin, and J.~Lu, ``Hamming distortion based secrecy systems: To foil
  the eavesdropper with finite shared key,'' \emph{IEEE Communications
  Letters}, vol.~19, no.~5, pp. 711--714, 2015.

\bibitem{klinc2009ldpc}
D.~Klinc, J.~Ha, S.~W. McLaughlin, J.~Barros, and B.-J. Kwak, ``{LDPC} codes
  for physical layer security,'' in \emph{IEEE Global Telecommunications
  Conference (GLOBECOM)}, 2009, pp. 5765--5770.

\bibitem{baldi2012coding}
M.~Baldi, M.~Bianchi, and F.~Chiaraluce, ``Coding with scrambling,
  concatenation, and harq for the awgn wire-tap channel: A security gap
  analysis,'' \emph{IEEE Trans. Inf. Forensics Security}, vol.~7, no.~3, pp.
  883--894, 2012.

\bibitem{harrison2011coding}
W.~K. Harrison, J.~Almeida, S.~W. McLaughlin, and J.~Barros, ``Coding for
  cryptographic security enhancement using stopping sets,'' \emph{IEEE Trans.
  Inf. Forensics Security}, vol.~6, no.~3, pp. 575--584, 2011.

\bibitem{thorpe2003low}
J.~Thorpe, ``Low-density parity-check {(LDPC)} codes constructed from
  protographs,'' \emph{IPN progress report}, vol.~42, no. 154, pp. 42--154,
  2003.

\bibitem{ten2003design}
S.~ten Brink and G.~Kramer, ``Design of repeat-accumulate codes for iterative
  detection and decoding,'' \emph{IEEE Transactions on Signal Processing},
  vol.~51, no.~11, pp. 2764--2772, 2003.

\end{thebibliography}

\end{document}